\documentclass[sigconf]{acmart}
\usepackage{tabularx}
\usepackage{booktabs}
\usepackage{multirow}
\usepackage{hyperref}
\usepackage{cleveref}
\usepackage{algorithm}
\usepackage{algorithmic}
\usepackage{setspace}
\usepackage{enumitem}
\usepackage{balance}

\AtBeginDocument{%
  \providecommand\BibTeX{{%
    \normalfont B\kern-0.5em{\scshape i\kern-0.25em b}\kern-0.8em\TeX}}}

\copyrightyear{2025}
\acmYear{2025}
\setcopyright{acmlicensed}\acmConference[KDD '25]{Proceedings of the 31st ACM SIGKDD Conference on Knowledge Discovery and Data Mining V.1}{August 3--7, 2025}{Toronto, ON, Canada}
\acmBooktitle{Proceedings of the 31st ACM SIGKDD Conference on Knowledge Discovery and Data Mining V.1 (KDD '25), August 3--7, 2025, Toronto, ON, Canada}
\acmDOI{10.1145/3690624.3709263}
\acmISBN{979-8-4007-1245-6/25/08}

\begin{document}
\newtheorem{thm}{Theorem} 
\newtheorem{defn}[thm]{Definition} 
\newtheorem{lem}[thm]{Lemma}

\title{MGS3: A Multi-Granularity Self-Supervised \\Code Search Framework}

\author{Rui Li}
\affiliation{%
  \institution{State Key Laboratory of Cognitive Intelligence, University of Science and Technology of China}
  \city{Hefei}
  \country{China}
}
\email{ruili2000@mail.ustc.edu.cn}

\author{Junfeng Kang}
\affiliation{%
  \institution{State Key Laboratory of Cognitive Intelligence, University of Science and Technology of China}
  \city{Hefei}
  \country{China}
}
\email{kangjf@mail.ustc.edu.cn}

\author{Qi Liu}
\authornote{Qi Liu is the corresponding author.}
\affiliation{%
  \institution{State Key Laboratory of Cognitive Intelligence, University of Science and Technology of China \& Institute of Artificial Intelligence, Hefei Comprehensive National Science Center}
  \city{Hefei}
  \country{China}
}
\email{qiliuql@ustc.edu.cn}

\author{Liyang He}
\affiliation{%
  \institution{State Key Laboratory of Cognitive Intelligence, University of Science and Technology of China}
  \city{Hefei}
  \country{China}
}
\email{heliyang@mail.ustc.edu.cn}

\author{Zheng Zhang}
\affiliation{%
  \institution{State Key Laboratory of Cognitive Intelligence, University of Science and Technology of China}
  \city{Hefei}
  \country{China}
}
\email{zhangzheng@mail.ustc.edu.cn}

\author{Yunhao Sha}
\affiliation{%
  \institution{State Key Laboratory of Cognitive Intelligence, University of Science and Technology of China}
  \city{Hefei}
  \country{China}
}
\email{percy@mail.ustc.edu.cn}

\author{Linbo Zhu}
\affiliation{%
  \institution{University of Science and Technology of China \& Institute of Artificial Intelligence, Hefei Comprehensive National Science Center}
  \city{Hefei}
  \country{China}
}
\email{lbzhu@iai.ustc.edu.cn}

\author{Zhenya Huang}
\affiliation{%
  \institution{State Key Laboratory of Cognitive Intelligence, University of Science and Technology of China \& Institute of Artificial Intelligence, Hefei Comprehensive National Science Center}
  \city{Hefei}
  \country{China}
}
\email{huangzhy@ustc.edu.cn}







\renewcommand{\shortauthors}{Rui Li et al.}

\begin{abstract}
In the pursuit of enhancing software reusability and developer productivity, code search has emerged as a key area, aimed at retrieving code snippets relevant to functionalities based on natural language queries. Despite significant progress in self-supervised code pre-training utilizing the vast amount of code data in repositories, existing methods have primarily focused on leveraging contrastive learning to align natural language with function-level code snippets. These studies have overlooked the abundance of fine-grained (such as block-level and statement-level) code snippets prevalent within the function-level code snippets, which results in suboptimal performance across all levels of granularity.
To address this problem, we first construct a multi-granularity code search dataset called \textbf{MGCodeSearchNet}, which contains 536K+ pairs of natural language and code snippets. Subsequently, we introduce a novel \textbf{M}ulti-\textbf{G}ranularity \textbf{S}elf-\textbf{S}upervised contrastive learning code \textbf{S}earch framework (\textbf{MGS$^{3}$}). First, MGS$^{3}$ features a Hierarchical Multi-Granularity Representation module (HMGR), which leverages syntactic structural relationships for hierarchical representation and aggregates fine-grained information into coarser-grained representations. Then, during the contrastive learning phase, we endeavor to construct positive samples of the same granularity for fine-grained code, and introduce in-function negative samples for fine-grained code.
Finally, we conduct extensive experiments on code search benchmarks across various granularities, demonstrating that the framework exhibits outstanding performance in code search tasks of multiple granularities. These experiments also showcase its model-agnostic nature and compatibility with existing pre-trained code representation models.
\end{abstract}





\maketitle

\title{MGS3: A Multi-Granularity Self-Supervised Code Search Framework}

\begin{figure}[t]
\includegraphics[width=0.47\textwidth]{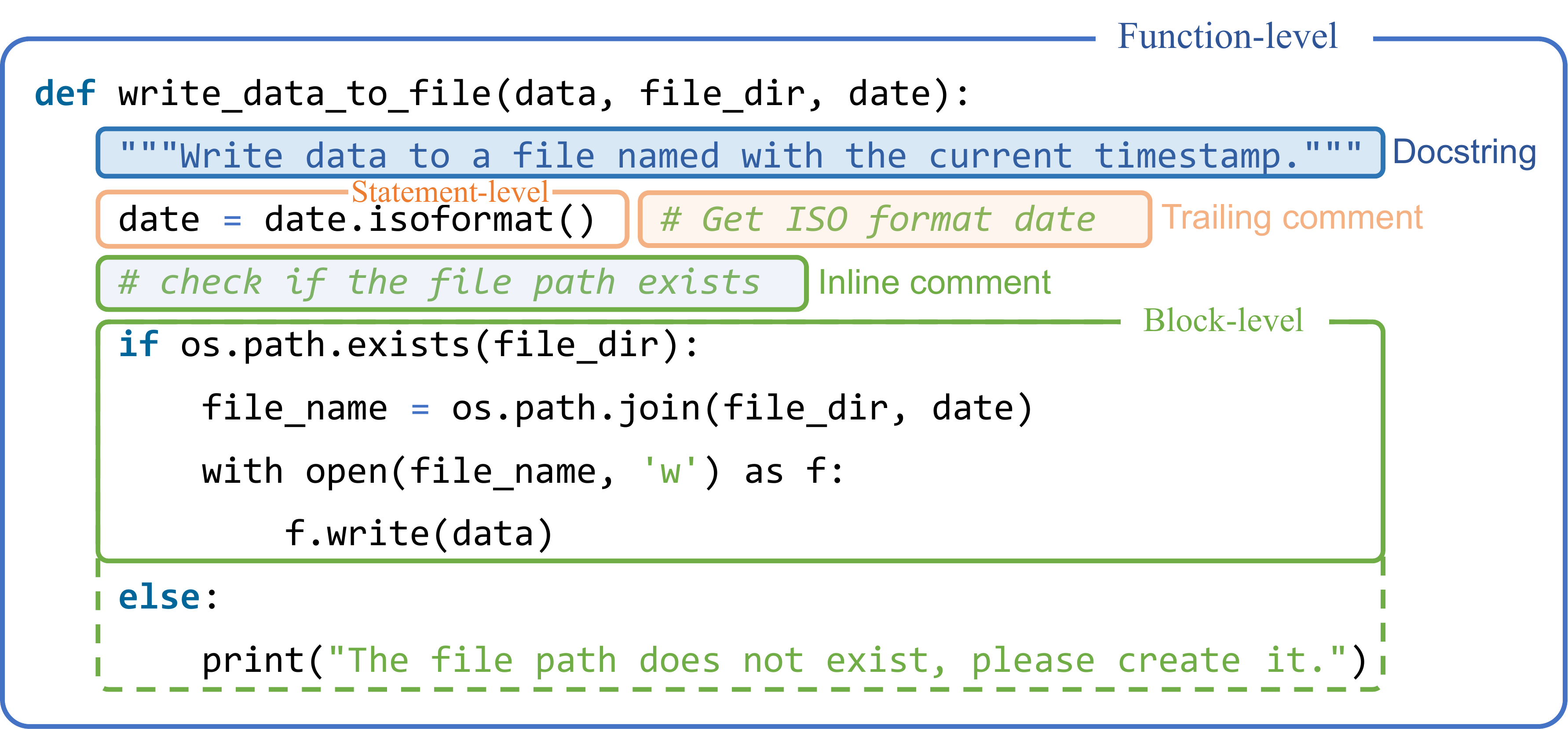}
\caption{Examples of functions in the repository, which encompass a variety granularities of code snippets, such as those at the function level, block level (e.g., \textit{if...}), and statement level (e.g., \textit{date=date.isoformat()}). Furthermore, the function has three types of natural language comments, including docstrings, inline comments, and trailing comments.}
\label{fig:demo}
\end{figure}

\section{Introduction}
Code search aims to retrieve functionally relevant code given a natural language query, facilitating software reuse and enhancing developer productivity. With the rapid expansion of online code repositories such as GitHub\footnote{\url{https://www.github.com}}, current research \cite{guo2022unixcoder, wang2023codet5+, li2024consider, li2024optimizing} predominantly focuses on leveraging the rich code data within these repositories for self-supervised code pretraining. Along this line, research work like CodeBERT \cite{feng2020codebert} have attempted to use the masked language modeling (MLM) pretraining task on large-scale code data to improve code representations. However, these methods focus on token-level pretraining tasks, which results in an inability to perform high-quality sequence encoding. To design training objectives more suitable for code search, several studies \cite{bui2021Corder, li2022coderetriever, li2024consider} have turned to contrastive learning for code pre-training in order to enhance the performance of code search.

In practical scenarios, there is often a need for code snippet searches of varying granularity  (i.e., statement-level, block-level, and function-level) to satisfy the diverse programming requirements of users. For instance, as shown in Figure 1, a user may wish to save a data file, in which case a function-level \textit{"write\_data\_to\_file"} would meet their needs. Alternatively, a user might simply need to convert the current date and time to the ISO format, and a brief statement of code such as \textit{"date=date.isoformat()"} would suffice for this purpose. 
However, prior research has predominantly focused on extracting \textbf{function-level} code snippets from repositories to construct alignment signals, overlooking the rich repository of fine-grained code snippets. As shown in \autoref{fig:demo}, within a function, there are statement-level code snippets \textit{"date=date.isoformat()"} and block-level code snippets: \textit{"if..."}.
This motivates us to explore more data-efficient methods for utilizing the diverse granularities of code snippets within repositories, aiming to achieve a \textbf{data-efficient code search across all levels of granularity}.

Through in-depth analysis, we have discovered that repositories contain a wide range of natural language comment signals within functions that can be aligned with code snippets at various granularities. For instance, as illustrated in Figure 1, there is an alignment between the inline comment \textit{"check if the file path exists"} and the block-level code snippet \textit{"if…"}, and between the trailing comment \textit{"Get ISO format date"} and the statement-level code snippet \textit{"date=date.isoformat()"}. This observation presents an opportunity to bridge this gap. Specifically, we extract natural language comments from functions in the code repository and pair them with code snippets at different granularities using coding standards and heuristic methods, thereby constructing a code search dataset called \textbf{MGCodeSearchNet}. MGCodeSearchNet comprises samples of diverse granularities, including 90.7K function-level pairs, 197K block-level pairs, and 258K statement-level pairs.

Moreover, enhancing multi-granularity code search with these natural language comments brings the following technical challenges.
\textbf{On the one hand}, there is syntactic structural information between multi-granularity code snippets. As shown in~\autoref{fig:demo}, a statement \textit{"f.write(data)"} exists within a \textit{"with..."} block. Therefore, leveraging these structural relationships to enhance representations across multiple granularities is crucial.
\textbf{On the other hand}, due to the absence of strict grammatical rules that define the scope of code referred to by comments, this leads to ambiguity in alignment during the pairing process. As shown in Figure 1, the inline comment \textit{"check if the file path exists"} could align with the subsequent \textit{"if..."} block, or it might align with the \textit{"if...else..."} block. This situation exceeds the scope of simple syntactic analysis and requires the integration of both the comment and the semantics of the code snippet for determination.

To tackle the remaining challenges, we introduce a novel \textbf{M}ulti-\textbf{G}ranularity \textbf{S}elf-\textbf{S}upervised contrastive learning code \textbf{S}earch framework (\textbf{MGS$^{3}$}) for code search models. To leverage structural information to enhance code snippet representations, we introduce a  \textbf{H}ierarchical \textbf{M}ulti-\textbf{G}ranularity \textbf{R}epresentation (\textbf{HMGR}) module that uses the syntactic structural relationships between different fine-grained code snippets for hierarchical representation, aggregating fine-grained representation information into coarse-grained representations for enhanced representation. 
To address the ambiguity in the alignment process, particularly when natural language comments need to be aligned with multiple candidate code snippets, we are inspired by \cite{khattab2020colbert} to adopt the \texttt{MaxSim} operators during the training process. 
In addition, for fine-grained code, we conduct negative sample mining in the repository to enhance the contrastive learning process of fine-grained code snippets by adding additional in-function negative samples.

We validate our framework's effectiveness through extensive experiments across various datasets featuring different granularities of code search, and further experimental results demonstrate that MGS$^{3}$ possesses good interpretability. 
It is worth mentioning that our proposed approach is model-agnostic and can be initialized using existing pre-trained code representation models.
In summary, our main contributions are summarized as follows:
\begin{itemize}[leftmargin=*]
\item We explore a novel perspective on self-supervised code search, attempting to extract supervision signals between code snippets of various granularities and natural language comments from widely available repositories. 
\item We introduce a multi-granularity code search dataset, \textbf{MGCodeSearchNet}, which pairs various types of natural language comments with code snippets of differing granularities within repositories using heuristic methods. 
\item We propose MGS$^{3}$, a multi-granularity self-supervised code search framework. MGS$^{3}$ enhances structural information through HMGR, improves the representation of code snippets, and enhances contrastive learning by mining negative samples.
\item We apply our framework to various pre-trained models and conduct experiments on a wide range of functional-level and snippet-level code search datasets. The results confirm that our method can improve the performance of existing pre-trained models and has good interpretability.
\end{itemize}

\section{Related Work}
\subsection{Code Search}
Due to the significant enhancement in model comprehension ability brought about by token-level pre-training tasks like Masked Language Modeling (MLM), CodeBERT \cite{feng2020codebert} and CuBERT \cite{Kanade2020cubert} have attempted to utilize the extensive programming language and natural language bimodal data in repositories for pre-training. GraphCodeBERT \cite{guo2021graphcodebert} attempts to introduce data flow graph signals on this basis to guide MLM pre-training, proving the importance of modeling data flows for code understanding. SynCoBERT \cite{wang2021syncobert} proposes a syntax-guided multimodal contrastive pre-training method that enhances the model's code representation capability by designing two new pre-training objectives: identifier prediction and AST edge prediction. 
UniXcoder \cite{guo2022unixcoder} introduces a unified cross-modal pre-training model designed for programming languages. This model leverages cross-modal content such as Abstract Syntax Trees (AST) and code comments to enhance code representation, thereby improving performance in code search tasks.

Recently, some works have attempted to enhance code search tasks using contrastive learning methods~\cite{he2023efficient, zhang2023fairlisa, he2024one}. ContraCode \cite{jain2021ContraCode} categorizes existing code transformation techniques into three types: code compression, identifier renaming, and canonicalization transformations. It employs these techniques for data augmentation and utilizes the objectives of contrastive learning to distinguish between similar and dissimilar code snippets. CodeRetriever \cite{li2022coderetriever} attempts to combine unimodal and bimodal contrastive learning to train function-level code search models.

However, these methods overlook supervisory signals other than the function-level. We use a heuristic approach to align code snippets at multiple granularities with natural language comments from the repository, thereby enhancing the performance of the model on different granularity code search tasks.

\subsection{Structured Code Representation}
Source code is distinct from natural language as it encompasses abundant structural and semantic information. Consequently, to obtain structured code representation, some early works \cite{iyer2016summarizing, ahmad2020transformer, wei2019code} treated source code as sequences or employed a Structure-Based Traversal (SBT) method \cite{hu2018deep} to flatten the Abstract Syntax Tree (AST), which allows the structural information within the program to enhance the representation of code. To better present the structure of the AST while maintaining the clarity of sequences, Hybrid-DeepCom \cite{hu2020HDeepcom} designed a novel structure-based traversal method to traverse the AST, which can explicitly recover the tree from the sequence generated by SBT and uses a hybrid attention component to merge lexical and syntactic information. CAST \cite{shi2021cast} proposed a method for hierarchical splitting and reconstruction of ASTs, aggregating the embeddings of the split subtrees to obtain the representation of the complete AST. 

Subsequently, many researchers \cite{leclair2020improved, allamanis2018learning, alon2018code2seq, shi2021cast} attempted to use structural information to guide the attention calculation process in Transformer networks. TPTrans \cite{ahmad2020transformer} explored two representative path encoding methods and integrated them into a unified Transformer attention module by modifying the position encoding to include path encoding and explored their interactions. HiT \cite{zhang2023HiT} enhances the sequential representation of code by integrating a global, statement-level hierarchy with a local, token-level hierarchy. PA-former \cite{chai2022pyramid} attempted to model the relationships between phrases, tokens, sub-tokens, and their mappings, converting the source code into a pyramid representation, and applying a pyramid attention mechanism to effectively aggregate features across different levels.

\subsection{Contrastive Learning Sample Construction}
With the development of deep learning~\cite{liu2019ekt, wang2022neuralcd}, contrastive learning plays a key role in both unsupervised and self-supervised learning. Previous research \cite{qu2021rocketqa, xiong2020ANCE, li2025foundation} has highlighted the critical importance of enhancing the quality of positive and negative samples within the contrastive learning framework. Consequently, many researchers have sought better sampling strategies to improve search performance \cite{zhang2021AR2, xiong2020ANCE}. In the realm of code search tasks, numerous studies have attempted to construct positive and negative samples for contrastive learning. For instance, Corder \cite{bui2021Corder} uses various semantic-preserving code editing techniques for code refactoring to create positive samples. BOOST \cite{ding2021boost} introduces a structure-guided code transformation algorithm to generate positive samples that are functionally identical but structurally different. Li et al. \cite{li2022RA} endeavors to construct representational-level positive samples that maintain semantic consistency without the need for additional data processing and training. Li et al. \cite{li2023rethinking} introduces the Soft-InfoNCE technique to ameliorate the issue of false negatives in code search tasks and to explicitly differentiate the potential relevance of negative samples.

The existing research has primarily focused on constructing positive and negative samples at the function level, without exploring the issue of constructing negative samples within fine-grained code snippets. We propose a multi-granularity code search framework that identifies suitable negative samples for different granularities of code snippets to enhance the contrastive learning process.

\section{The Dataset: MGCodeSearchNet}
\label{sec:datacf}
\subsection{Data Collection}
To gather and align code snippets of varying granularity from a wide range of open-source repositories, we leverage the existing CodeSearchNet corpus. The CodeSearchNet corpus \cite{husain2019codesearchnet} comprises an extensive collection of publicly available code data harvested from non-forked open-source GitHub repositories, all of which have licenses that are publicly accessible. It consists of 2.1 million pairs of data (functions paired with docstring) in six popular programming languages: Go, Java, JavaScript, PHP, Python, and Ruby. As described in \cite{husain2019codesearchnet}, the docstring for the code is extracted from the function header comments. We adhere to the approach of \cite{guo2021graphcodebert} to filter out examples of low quality, resulting in 90.7K function-level code snippets.

To extract supervisory signals of varying granularity, we initially parse the functions within the codebase using the tree-sitter\footnote{\url{https://tree-sitter.github.io}} library. Subsequently, we employ a heuristic approach based on programming conventions and code structure to align natural language comments with code snippets. Specifically, we first obtain natural language comments through parsing and then classify them according to \autoref{fig:demo}. Depending on the type, we employ different methods for alignment:
\begin{itemize}[leftmargin=*]
\item \textbf{Docstrings}: we follow the practices established in prior research, aligning them with the entire function body. 
\item \textbf{Inline comments}: our focus is primarily on extracting snippets of code that contain control flows. We attempt to collect the subsequent code snippet based on the comment information; however, this heuristic method has a flaw that can lead to ambiguous alignment information. For instance, in \autoref{fig:demo}, it is unclear whether the natural language comment \textit{"check if the file path exists"} should align with the \textit{"if..."} block or the \textit{"if...else..."} block. Our approach to handling this involves establishing a \textbf{one-to-many} alignment relationship between natural language comments and different code snippets. We then utilize semantic knowledge during the modeling process to make determinations. 
\item \textbf{Trailing comments}: we directly align these natural language comments with the code snippets of the current line. 
\end{itemize}
Using the aforementioned heuristic alignment method, we can establish alignment relationships between various types of natural language comments and code snippets of multiple granularities.

\begin{table}[tb]
\caption{The statistical results of our multi-granularity code search dataset.}
\scalebox{0.94}{
\begin{tabular}{cllllll}
\hline
   Granularity & \multicolumn{1}{c}{Ruby} & \multicolumn{1}{c}{JS} & \multicolumn{1}{c}{Go} & \multicolumn{1}{c}{Python} & \multicolumn{1}{c}{Java} & \multicolumn{1}{c}{PHP} \\ \hline
Function-level & 2.5K                      & 5.8K                    & 16.7K                    & 25.2K                        & 16.4K                      & 24.1K                     \\
Block-level & 1.7K                      & 18.6K                    & 20.7K                    & 55.9K                        & 47.6K                      & 52.9K                     \\
Statement-level & 8.5K                      & 20.9K                    & 45.8K                    & 84.5K                        & 24.8K                      & 73.6K                     \\ \hline
\end{tabular}
}
\label{tab:dataset}
\end{table}

\begin{figure*}[htbp]
\begin{center}
\includegraphics[width=1.0\linewidth]{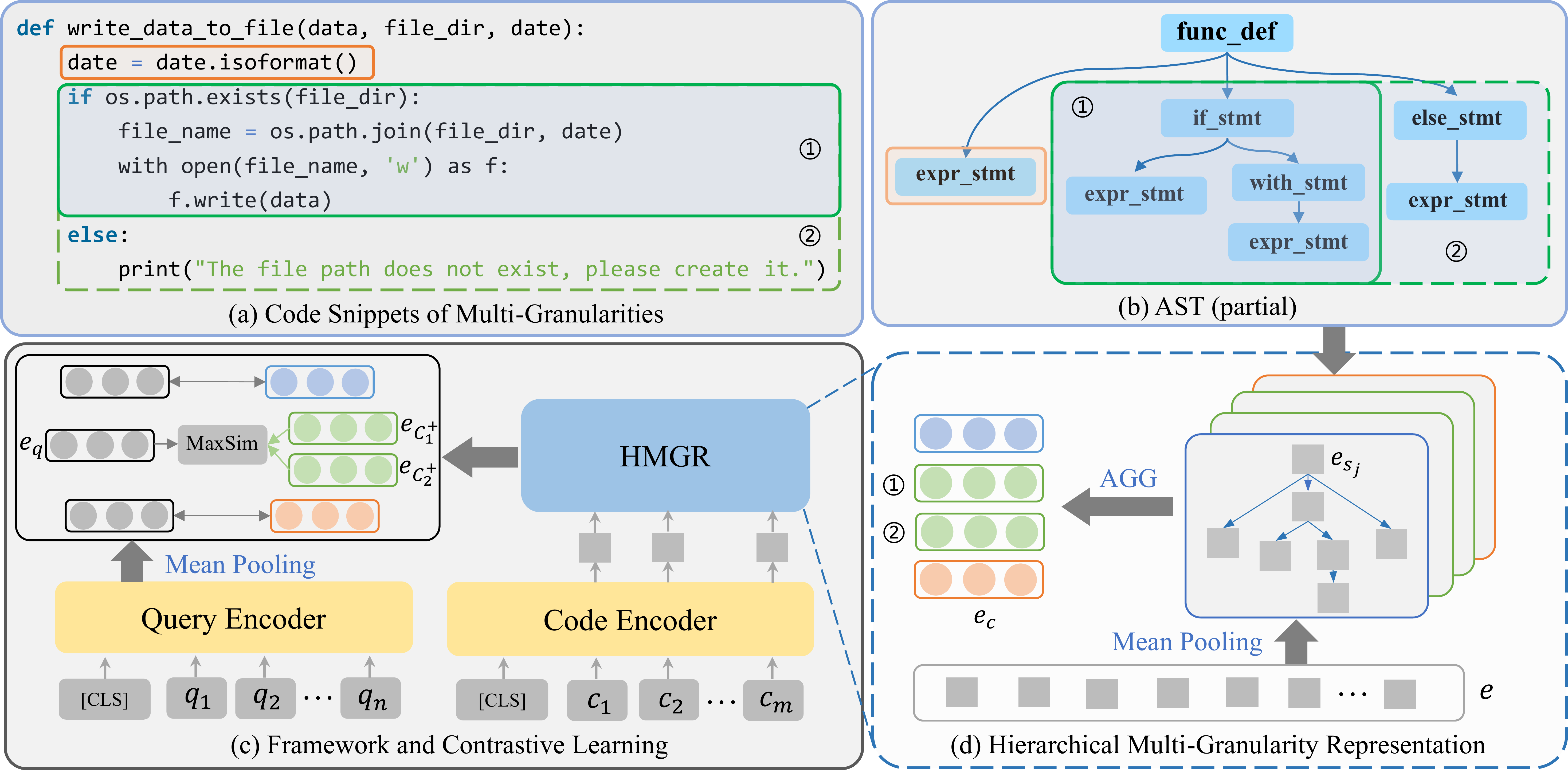}
\end{center}
\caption{The overview of MGS$^{3}$. (a) Different granularity code snippets. (b) Depicts the abstract syntax tree (AST) corresponding to the code snippets, retaining only the statement and block nodes. (c) Illustrates the bi-encoder model using MaxSim for contrastive learning among code snippets. (d) The hierarchical aggregation in HMGR guided by AST.}
\label{fig:framework}
\end{figure*}

\subsection{Data Filtering}
Due to the presence of substantial noise generate by programmers during the coding process within function comments, it may seriously damage the quality of the dataset. We attempt to utilize regular expressions to filter the extracted natural language comments. Specifically, we apply filtering in the following scenarios:
\begin{itemize}[leftmargin=*]
    \item Comments that include URL information (e.g., "https://...");
    \item Comments starting with special terms (e.g., "TODO");
    \item Comments used for automated code reviews (e.g., "Linter...");
    \item Comments that are shorter than four tokens.
\end{itemize}
In addition, by calculating the TF-IDF~\cite{ramos2003using} scores of the current code snippet and the global code snippet, we will also filter out pairs that overly rely on the global code instead of the current fine-grained code snippet. The resulting collect MGCodeSearchNet includes 536K+ (query, code) pairs from 35K+ projects. The statistical information of the dataset is shown in \autoref{tab:dataset}. We present more statistical results of MGCodeSearchNet in \autoref{sec:datadetails}.

\section{Preliminaries}   
\subsection{Code Search}
\label{sec:codesearch}
Code search is a technique for retrieving code snippets that are most relevant to a given query. During training, the model is trained by maximizing the similarity between the queries and the corresponding code. In the given space of (query, code snippet) pairs $(Q, C)$, we denote $(q,c) \in (Q, C)$ as a pair. Here $q = \{q_1, q_2,...,q_{n}\}$ represents a query composed of $n$ tokens, and $c = \{c_1, c_2,…c_{m}\}$ represents a code snippet sequence composed of $m$ tokens. We employ a query encoder $E_Q$ and a code encoder $E_C$ to encode the query $q$ and the code $c$ respectively. Subsequently, we train $E_Q$ and $E_C$ (in practice, parameters are typically shared between $E_Q$ and $E_C$) to satisfy the following conditions for each query code pair $(q,c)$:
\begin{equation}
    \forall q \in Q, \max_{c \in C} f\left(q,c\right),
\end{equation}
where $f(\cdot,\cdot)$ is used to calculate the matching score between a query and a code snippet. We use dot product to compute the similarity between query-code pairs. we employ the InfoNCE loss \cite{oord2018infonce} for contrastive learning. Additionally, for efficient training, we adopt the in-batch negative strategy to construct negative sample pairs, aiming to minimize the similarity between them. Specifically, For a given batch of data, each query $q$ can be paired with a positive sample code snippet $c^+$, as well as with $N - 1$ other code snippets $\{c^{-}_{1}, c^{-}_2, \ldots, c^{-}_{N-1}\}$ to create $N - 1$ negative sample pairs, where $N$ is the batch size. The InfoNCE loss can be described as follows:
\begin{equation}
    \mathcal{L}=-\mathbb{E}\left[\log \frac{e^{f\left(q, c^+\right)/\tau}}{e^{f\left(q, c^+\right)/\tau}+\sum_{j=1}^{N-1}e^{f(q, c^{-}_{j})/\tau}}\right],
\end{equation}
where $\tau$ represents the temperature parameter in the contrastive learning loss function, which can generally be used to adjust the level of attention that the contrastive learning loss function pays to negative samples \cite{zhang2024enhancing, he2020moco, zhao2023ensemble}.

\section{THE MGS$^{3}$ FRAMEWORK}
\subsection{Framework Overview}
As depicted in \autoref{fig:framework}, to address the challenges mentioned earlier, we propose a \textbf{M}ulti-\textbf{G}rained \textbf{S}elf-\textbf{S}upervised Code \textbf{S}earch Framework (MGS$^{3}$) as a solution. Specifically, we use the existing code representation model as the backbone and MGCodeSearchNet as the training dataset.  To leverage the structural information across different granularities for representing code snippets, we introduce a Hierarchical Multi-Grained Representation (HMGR) module in \autoref{sec:structuredMG}. After aligning natural language comments with code snippets and obtaining representations of code at various granularities, we discuss the construction of positive and negative samples within the contrastive learning objective in \autoref{sec:MGRA}.

\subsection{Hierarchical Multi-Granularity Representation}
\label{sec:structuredMG}
Due to the need for representing code snippets at multiple granularities and the existence of structured relationships between different granularities within a function, we propose a Hierarchical Multi-Granularity Representation (HMGR) module. We consider code snippets at the statement-level granularity as the smallest unit of representation. For other coarser-grained code snippets, we attempt to aggregate their representations using hierarchical sub-node information within the code snippet. The difference from the previous method of using structured information from code is that we do not focus on more fine-grained token-level information, which reduces the complexity of the structure and provides a higher level of semantic guidance.

Since fine-grained code snippets contain contextual information within a function, extracting such code snippets directly from the function and aligning them with a query would result in the loss of contextual information. Therefore, regardless of the granularity of the code snippet being encoded, we treat the entire function as the input to the model and employ different extraction methods for different granularities of code snippets. Specifically, as shown in \autoref{fig:framework} (c), we first encode the input code of the current code snippet using a code encoder model:
\begin{equation}
    e =  E_C([CLS], c_1, c_2, ..., c_n).
\end{equation}
By encoding the function, we ultimately obtain the embeddings of all tokens in the function, i.e., $e = \{e_{cls}, e_1, e_2, ..., e_n\}$. After obtaining representations for all tokens in the function, we retrieve the representation of each token and use mean pooling to obtain the representation of all statement-level code snippets:
\begin{equation}
    e_{S_j} = \frac{1}{|S_j|}\sum_{e_i\in S_j}e_i,
\end{equation}
where $S$ represents the collection of code snippets represented at the statement level, that is, $S = \{S_{j}\}_{j=1}^{|S|}$.
For function-level and block-level code snippets, hierarchical relationships exist within these code snippets. Taking block-level code snippets as an example, typically, a block encompasses multiple statement-level code snippets and other block-level code snippets. Moreover, there are hierarchical relationships among these code snippets (e.g., in \autoref{fig:framework} (a), the \textit{"with..."} loop block includes the \textit{"f.write(data)"} statement). 

Therefore, during the modeling process of the current block-level code snippet, we consider aggregating the representations of its child nodes in a hierarchical manner. Specifically, as shown in \autoref{fig:framework} (b), we first parse the function to obtain the abstract syntax tree (AST) and extract the subtree $T$ corresponding to the current code snippet $c$. Then, as shown in \autoref{fig:framework} (d), we utilize the AST to guide the aggregation of the statement-level representations:
\begin{equation}
    \widehat{e}_c = AGG(e_c, \{e_v: v \in V(T)\text{ and }parent(v) = c\}),
\end{equation}
where \(V(T)\) signifies the set of nodes in the syntactic tree $T$. The function 'parent' is used to identify the parent node of a given node \(v\) in tree \(T\). The \(AGG\) function represents an aggregation mechanism that combines the representation of the root node of the current code snippet with the information from all its child nodes:
\begin{equation}
    AGG(e_c, S) = \text{LayerNorm}(e_c + W * \frac{1}{\left|S\right|} \sum_{v \in S}{e_v}),
\end{equation}
where \(W\) represents trainable parameters.

Due to code snippets in the codebase can be constructed during the offline phase, there is actually only a need to represent the query during the online stage. Consequently, a significant advantage of our proposed HMGR is that it does not increase the online latency.

\subsection{Multi-Granularity Alignment}
\label{sec:MGRA}
Our objective is to leverage the alignment information between comments and code snippets for contrastive learning. The essence of the contrastive learning lies in the construction of appropriate positive and negative samples. Therefore, we attempt to construct contrastive samples for code snippets of varying granularities.

\subsubsection{\textbf{Positive Sample Construction}}
In \autoref{sec:datacf}, we constructed positive sample pairs at various granularities through analytical construction. However, the use of heuristic algorithms for alignment can result in one natural language comment corresponding to multiple candidate block-level code snippets. Consequently, we aim to utilize the semantic information between natural language comments and candidate code snippets to select appropriate positive samples for contrastive learning.

Inspired by prior research \cite{khattab2020colbert}, we employed a \texttt{MaxSim} operator for alignment. Specifically, as shown in the lower part of \autoref{fig:framework} (c), for the natural language comment $q$ and the potentially corresponding code snippet sets $C^{+} = \{C_{1}^{+}, C_{2}^{+}, \cdots, C_{n}^{+}\}$, we calculate similarity scores using their embeddings. During contrastive learning, we select the code snippet with the highest score to align as the positive sample with the natural language comment:
\begin{equation}
    f(q, C^{+}) = \max(e^{\top}_{q} e_{C_1^{+}}, \ldots, e^{\top}_{q} e_{C_n^{+}}),
\end{equation}
where $e_q$ is the encoding of the query. We obtain it by encoding the tokens of the query input into a model and then applying mean pooling. Additionally, we use HMGR from \autoref{sec:structuredMG} to obtain the code representation of $e_{C^{+}_{i}}$.


\subsubsection{\textbf{Negative Sample Construction}}
Since our constructed dataset contains multiple granularities, it is necessary to select appropriate negative samples for each granularity for contrastive learning. Previous research \cite{li2023rethinking, qu2021rocketqa} has already demonstrated that selecting as difficult negative samples as possible can facilitate the process of contrastive learning. Therefore, we choose negative samples of the same granularity for each code snippet during contrastive learning. For all granularities of code snippets, we can try to collect negative samples from other functions. Additionally, for block-level and statement-level code snippet samples, we have observed that these fine-grained code snippets allow for another source of negative samples — negative samples of corresponding granularity within the same function. 

Specifically, MGS$^{3}$ extends the negative sample set $c^-$ to include other code snippets of the same granularity within the same function. We consider that these internal negative samples from the same function can be considered more challenging than random samples from blocks and statements of other functions because they share similar function context information. This requires the model to have the ability to distinguish specific functionalities within the same function context. We summarize the full negative sample construction process in \autoref{sec:negalgo}.

\subsection{Training Objective}
MGS$^{3}$ uses three levels of supervision signals for contrastive learning training from natural language comments to code snippets. As mentioned in \autoref{sec:codesearch}, we utilize the InfoNCE loss function for pre-training contrastive learning between natural language comments and code snippets at different granularities. We denote the loss function at the function level as $\mathcal{L}_{f}$, at the block level as $\mathcal{L}_{b}$, and at the statement level as $\mathcal{L}_{s}$. Therefore, our final training objective $\mathcal{L}_{MG}$ is: 
\begin{equation}
    \mathcal{L}_{MG}= \mathcal{L}_{f}+\alpha \mathcal{L}_{b}+\beta \mathcal{L}_{s}, \label{eq:loss}
\end{equation}
where $\alpha$ and $\beta$ are hyper-parameters to balance different granularity objective losses. The training goal is to minimize the integrated loss with respect to the model parameters. 

\begin{table*}[htb]
\caption{Under the zero-shot setting, we compare the original results of the pre-trained model with the performance using MGS$^{3}$ across various granularities benchmarks. We get the CodeT5+'s result by using the released checkpoint. Other results of compared models are reported by previous papers. All experiments meet the p<0.01 significance threshold.}
\scalebox{1.05}{
\begin{tabular}{cccccccccc}
\hline
\multirow{2}{*}{Granularity} & \multirow{2}{*}{Dataset} & \multicolumn{2}{c}{CodeBERT} & \multicolumn{2}{c}{GraphCodeBERT} & \multicolumn{2}{c}{UniXCoder} & \multicolumn{2}{c}{CodeT5+} \\ \cline{3-10} 
                             &                          & Original      & w/ MGS$^{3}$      & Original         & w/ MGS$^{3}$        & Original       & w/ MGS$^{3}$      & Original      & w/ MGS$^{3}$     \\ \hline
\multirow{3}{*}{Function-level}    & Adv                      & 0.5             & \textbf{25.6}            & 0.5               & \textbf{26.3}              & 2.4              & \textbf{27.8}            & 28.5               & \textbf{29.7}           \\
                             & CoSQA                    & 0.9             & \textbf{36.6}            & 0.8                & \textbf{37.7}              & 9.5              & \textbf{39.9}            & 39.4             & \textbf{41.5}           \\ 
                             & XLCoST-FL                    & 0.7             & \textbf{26.0}            & 0.8                & \textbf{26.4}              & 2.2              & \textbf{28.3}            & 30.2             & \textbf{33.7}           \\ \hline                             
\multirow{3}{*}{Block-level}       & SO-DS                    & 0.4             & \textbf{16.4}            & 0.6                & \textbf{16.7}              & 0.5              & \textbf{16.3}            & 9.2             & \textbf{18.9}           \\
                             & StaQC                    & 0.6             & \textbf{14.7}            & 0.5                & \textbf{14.8}              & 0.7              & \textbf{14.5}           & 8.3             & \textbf{16.5}           \\
                             & XLCoST-BL                    & 0.9             & \textbf{20.7}            & 1.2                & \textbf{21.0}              & 0.8              & \textbf{21.4}            & 13.4             & \textbf{25.8}           \\ \hline                               
Statement-level                    & CoNaLa                   & 2.0             & \textbf{12.8}            & 2.5                & \textbf{13.1}              & 2.7              & \textbf{12.6}            & 10.2             & \textbf{15.0}           \\ \hline
\end{tabular}
}
\label{tab:main1}
\end{table*}

\section{Experiment}
We conducted comprehensive experiments to answer the following research questions:
\begin{itemize}[leftmargin=*]
    \item \textbf{RQ1}: How does the MGS$^{3}$ perform across various granularities of code search tasks?
    \item \textbf{RQ2}: What is the adaptability of MGS$^{3}$ when fine-tuned for tasks involving multiple granularities of code search?
    \item \textbf{RQ3}: What are the roles of various modules in the MGS$^{3}$?
    \item \textbf{RQ4}: How interpretable is our proposed MGS$^{3}$?
\end{itemize}

\subsection{Benchmark Datasets}
To validate the performance of MGS$^{3}$ on code search tasks of different granularities, we evaluated it on several existing code search benchmarks of varying granularities:
\begin{itemize}[leftmargin=*]
    \item \textbf{Function level}: We select function-level evaluation benchmarks that are widely used in related work, including CodeSearchNet (CSN) \cite{husain2019codesearchnet, guo2021graphcodebert}, Adv \cite{lu2021codexglue}, CoSQA \cite{huang2021cosqa}, and XLCoST-FL \cite{zhu2022xlcost}. These datasets come from GitHub repositories, the StackOverflow\footnote{\url{https://www.stackoverflow.com}} community, and the GeeksForGeeks\footnote{\url{https://www.geeksforgeeks.org}}. 
    \item \textbf{Block level}: We choose SO-DS \cite{heyman2020sods} and StaQC \cite{yao2018staqc} and XLCoST-BL as block-level evaluation benchmarks, where SO-DS, StaQC are from Stackoverflow and XLCoST-BL is from GeeksForGeeks. 
    \item \textbf{Statement level}: CoNaLa \cite{yin2018conala} collects question titles and replies from StackOverflow to construct a code search dataset.
\end{itemize}
We have summarized the detailed statistics of these benchmarks and dataset information in \autoref{sec:benchdetails}.

\subsection{Comparison Methods and Metrics}
To validate that our propose MGS$^{3}$ framework can enhance the multi-granularity code search capabilities of existing pre-trained code representation models, we apply it to several models:
\begin{itemize}[leftmargin=*]
    \item \textbf{CodeBERT} \cite{feng2020codebert} is a bimodal pre-trained model that is pre-trained through two tasks: Masked Language Modeling (MLM) and Replaced Token Detection (RTD). 
    \item \textbf{GraphCodeBERT} \cite{guo2021graphcodebert} proposes two structure-based pre-training tasks (data flow edge prediction and node alignment) to enhance code representation.
    \item \textbf{UniXcoder} \cite{guo2022unixcoder} proposes to enhance code representation using cross-modal content such as AST and code comments.
    \item \textbf{CodeT5+} \cite{wang2023codet5+} uses a mix of pre-training objectives (span denoising, contrastive learning, etc.) for pre-training on monolingual and bilingual multi-language code corpora.
\end{itemize}
Following the previous research work, we use Mean Reciprocal Rank (MRR) \cite{hull1999mrr} as the evaluation metric on all benchmarks.
\begin{equation}
    MRR=\frac{1}{N}\sum_{i=1}^{N}{\frac{1}{rank_i}},
\end{equation}
where $rank_i$ is the rank of the correct code snippet related to the i-th query.

\subsection{Implementation Details}
 Our MGS$^{3}$ framework is implemented in PyTorch\footnote{Our code and dataset are available at \url{https://github.com/smsquirrel/MGS3}.}. We initialize the model using pre-trained code representation models, and for all models, we map the final output dimensions to 768. We employ the AdamW optimizer \cite{Loshchilov2017AdamW}, experimenting with learning rates set to \{1e-5, 2e-5, 5e-5\}. The batch size is empirically set to 256. We select the number of training epochs from the set \{20, 40, 60\}, attempting to use an early stopping strategy. The maximum sequence lengths for the text and code are set to 128 and 320, respectively. We select the temperature $\tau$ in contrastive learning from the set \{0.05, 0.1, 0.2\}. 
We set the hyperparameters in \autoref{eq:loss} as $\alpha=1$ and $\beta=0.6$. Detailed parameter sensitivity experiments are in Appendix~\ref{sec:paraSensty}.
The experiments described in this paper are conducted with three random seeds: 0, 1, and 2, and we will report the average results in the paper. All experiments meet the p<0.01 significance threshold. All experiments are performed on a Linux server equipped with four 2.30GHz Intel Xeon Gold 5218 CPUs and two Tesla A100 GPUs.

\begin{table*}[htb]
\caption{In benchmarks of code search at different granularities, we compare the fine-tuning performance of pre-trained code representation models with the performance under a fine-tuning setting using MGS$^{3}$.}
\begin{tabular}{cccccccccc}
\hline
\multirow{2}{*}{Granularity}    & \multirow{2}{*}{Dataset} & \multicolumn{2}{c}{CodeBERT} & \multicolumn{2}{c}{GraphCodeBERT} & \multicolumn{2}{c}{UniXCoder} & \multicolumn{2}{c}{CodeT5+} \\ \cline{3-10} 
                                &                          & Fine-tuning  & w/ MGS$^{3}$  & Fine-tuning    & w/ MGS$^{3}$     & Fine-tuning  & w/ MGS$^{3}$   & Fine-tuning & w/ MGS$^{3}$  \\ \hline
\multirow{4}{*}{Function-level} & CSN                      & 69.3         & \textbf{71.4} & 71.3           & \textbf{73.2}    & 74.4         & \textbf{75.7}  & 74.6        & \textbf{75.6} \\ & Adv                      & 27.2         & \textbf{29.4} & 35.2           & \textbf{37.2}    & 41.3         & \textbf{42.5}  & 43.3        & \textbf{44.7} \\
                                & CoSQA                    & 64.7         & \textbf{66.0} & 67.5           & \textbf{68.8}    & 70.1         & \textbf{71.3}  & 72.7        & \textbf{73.9} \\
                                & XLCoST-FL                & 58.0         & \textbf{60.7} & 59.4           & \textbf{62.2}    & 59.1         & \textbf{61.9}  & 63.3        & \textbf{65.1} \\
                                 \hline
\multirow{3}{*}{Block-level}    & SO-DS                    & 23.1         & \textbf{27.3} & 25.3           & \textbf{27.9}    & 23.6         & \textbf{27.3}  & 26.1        & \textbf{29.5} \\
                                & StaQC                    & 23.4         & \textbf{27.6} & 23.8           & \textbf{28.7}    & 23.1         & \textbf{28.2}  & 25.7        & \textbf{29.4} \\
                                & XLCoST-BL                & 30.9         & \textbf{38.9} & 31.6           & \textbf{39.5}    & 30.2         & \textbf{38.0}  & 35.8        & \textbf{40.0} \\ \hline
Statement-level                 & CoNaLa                   & 20.9         & \textbf{25.0} & 23.5           & \textbf{26.9}    & 20.4         & \textbf{25.1}  & 22.8        & \textbf{27.1} \\ \hline
\end{tabular}
\vspace{-5pt}
\label{tab:main2}
\end{table*}

\subsection{Zero-Shot Performance (RQ1)}

First, to validate the performance of MGS$^{3}$ on various granularity code search tasks, we apply MGS$^{3}$ to multiple pre-trained code representation models and conduct contrastive learning pre-training on the MGCodeSearchNet dataset. 
Subsequently, we evaluate the models using various granularity code search benchmarks. \autoref{tab:main1} presents the zero-shot performance of different methods. It is worth noting that since the MGCodeSearchNet data is constructed based on CodeSearchNet, we do not show the results on CodeSearchNet under zero-shot conditions. By observation, we find that applying MGS$^{3}$ on pre-trained code representation models leads to performance improvements on all benchmarks. 
Furthermore, we observe that code representation models trained with token-level pretraining tasks perform poorly on all benchmarks, but achieve significant performance improvements after applying MGS$^{3}$. This demonstrates the effectiveness of our model in enhancing the performance of pretrained code representation models on code search tasks. 
Due to the inclusion of function-level code snippet contrastive learning during the pre-training phase, CodeT5+ exhibits better performance in function-level granularity benchmarks. However, it performs poorly in other granularity search benchmarks. After applying the MGS$^{3}$ framework, performance improvements are observed across all benchmarks, particularly in block-level and statement-level search benchmarks. This further validates the effectiveness of the MGS$^{3}$ framework.

\begin{figure}[htbp]
\begin{center}
\includegraphics[width=1.0\linewidth]{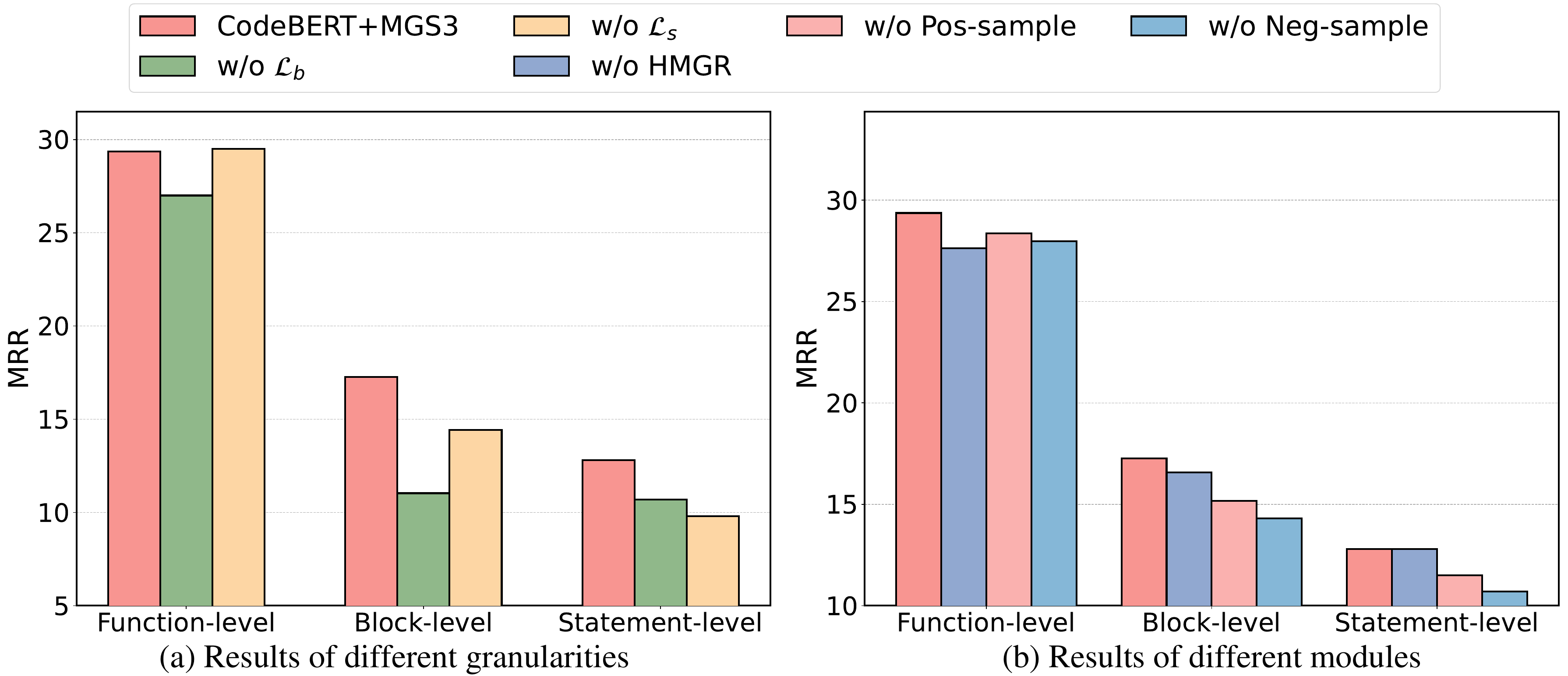}
\end{center}
\caption{Results of ablation study on different granularities and modules. Full results can be found in Appendix~\ref{sec:ablation}.}
\label{fig:ablation}
\end{figure}

\subsection{ Fine-Tuning Performance (RQ2)}
To validate the fine-tuning performance of the MGS$^{3}$ framework, we conduct fine-tuning on the training sets of various downstream benchmarks subsequent to the pre-training phase. We then evaluate the model's performance on the corresponding test sets. The experimental results are presented in \autoref{tab:main2}. Observation of these results reveals that, compared to other baselines, the application of the MGS$^{3}$ framework significantly enhances performance on the CodeSearchNet dataset. This suggests that our method effectively leverages a broader range of supervisory signals within repositories to improve model performance. Furthermore, our approach achieves the best performance across various granularity search benchmarks, demonstrating that models pre-trained with the MGS$^{3}$ framework can be rapidly adapted to downstream code search tasks of different granularities. 
Further observation shows that our model exhibits significantly better performance on the XLCoST-BL dataset, which may be attributed to the close similarity in data distribution between XLCoST-BL and our training dataset.
It is also observed that our method demonstrates a more pronounced improvement on smaller training sets (such as CoNaLa), suggesting that our approach holds substantial potential for application in scenarios characterized by data sparsity.

\subsection{Ablation Study (RQ3)}
To examine the effectiveness of incorporating alignment signals of varying granularities and the role of each module within MGS$^{3}$, we conduct a series of ablation studies. We initialize our framework using the CodeBERT model and evaluate its performance across code search tasks of multiple granularities.
\subsubsection{\textbf{Effectiveness of Different Granularities}}
To investigate the impact of alignment signals at different granularities on the model, we conduct an ablation study. Specifically, we introduce the conditions "w/o $\mathcal{L}_{b}$" and "w/o $\mathcal{L}_{s}$," which exclude the loss terms for block-level and statement-level alignment signals, respectively, from \autoref{eq:loss}. The results, as shown in \autoref{fig:ablation} (a), indicate a noticeable decline in model performance on code search tasks corresponding to these granularities. This suggests that adding supervisory signals at specific granularities can effectively aid the code search model in adapting to code search tasks at those levels. Furthermore, the removal of $\mathcal{L}_{b}$ results in reduced performance on code search tasks at other granularities, implying that block-level alignment signals can enhance the model's understanding of code snippets at various granularities. Conversely, the elimination of $\mathcal{L}_{s}$ appears to improve performance on function-level code search tasks. We consider that this is due to significant differences between statement-level and function-level code snippets.

\begin{figure*}[htbp]
\begin{center}
\includegraphics[width=1.0\linewidth]{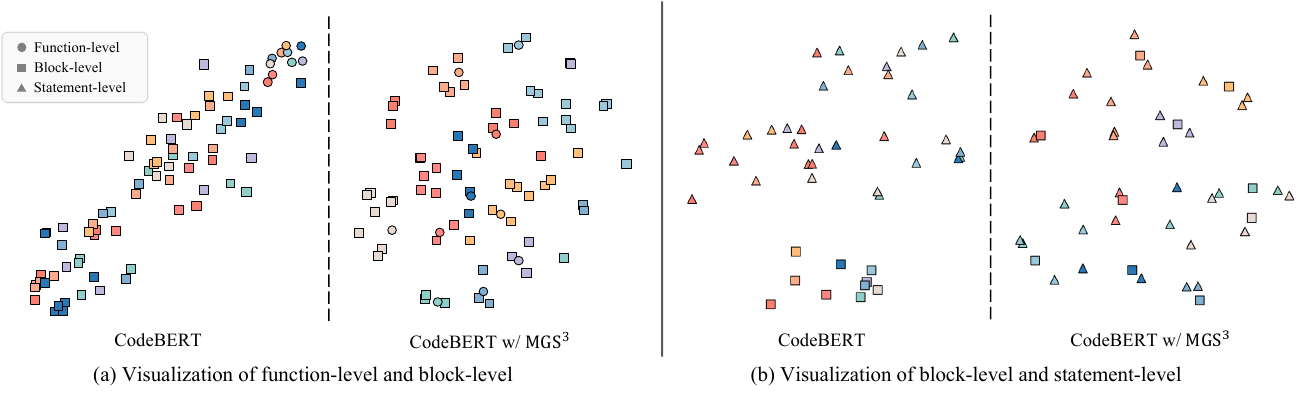}
\end{center}
\caption{Visualization of code snippets represented in multiple granularities. Code snippets with the same functionality are identified with the same color, while code snippets of different granularities are identified with different shapes.}
\label{fig:MGVisual}
\end{figure*}

\begin{figure}[htbp]
\begin{center}
\includegraphics[width=1.0\linewidth]{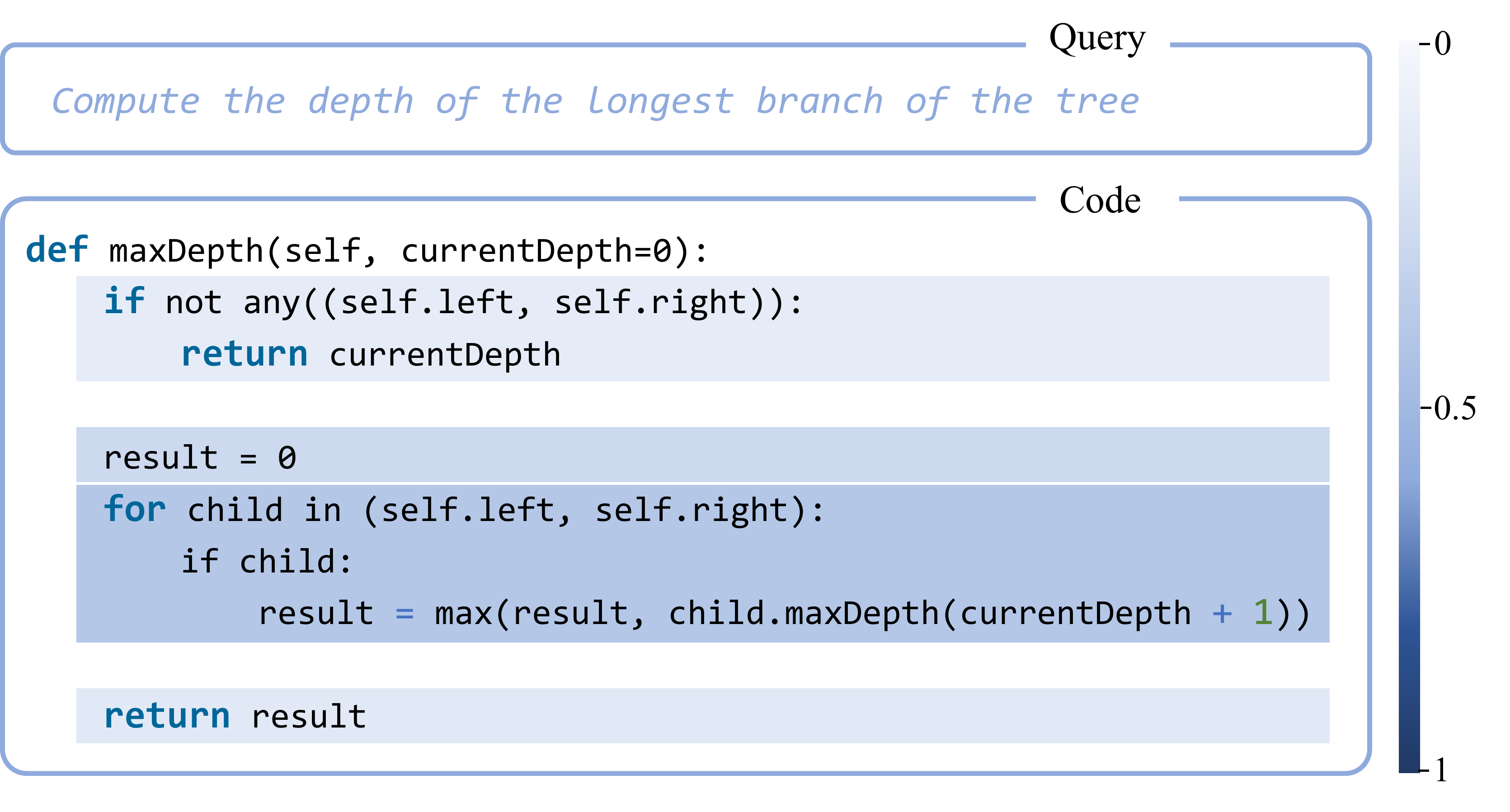}
\end{center}
\caption{Visualize the importance of final search scores for different granularity code snippets.}
\label{fig:casestudy}
\end{figure}

\subsubsection{\textbf{Effectiveness of Different Modules}}
To validate the contribution of each module within the MGS$^{3}$, We denote "w/o HMGR" to indicate the removal of the HMGR module and instead use mean pooling to obtain representations of code snippets at different granularities. "w/o Pos-sample" indicates the removal of alignment and instead aligning with the nearest largest block in natural language comments. "w/o Neg-sample" indicates the removal of in-function negative samples and only using in-batch negative samples for contrastive learning. 

The final experimental results are shown in \autoref{fig:ablation} (b), removing any module will lead to a decrease in performance on downstream tasks, confirming the effectiveness of our proposed modules. It is worth noting that we observe a decrease in performance on code search tasks at all granularities when removing the HMGR module, especially at the block-level and function-level. This suggests that the HMGR module helps models utilize structured information between different granularity code snippets. 
Additionally, we notice that "w/o Pos-sample" significantly reduces the performance of block-level code search tasks. This implies that this module contributes to more accurate and effective block-level alignment and proves that aligning natural language comments with code snippets solely through heuristic methods may result in suboptimal performance. 
Finally, we observe that "w/o Neg-sample" leads to a decrease in performance for both block-level and statement-level code search tasks. This indicates that in-function negative samples can increase the difficulty for fine-grained code snippet negatives and provide more diverse sources of negative samples, resulting in better contrastive learning outcomes.

\subsection{Visualization (RQ4)}
\subsubsection{\textbf{Representation Distribution of Code Snippets with Different Granularities.}}
In this section, we aim to visually demonstrate the capabilities of our framework. We first randomly selected 10 functions from the XLCoST dataset and parsed them to obtain the corresponding block-level code snippets, as well as collected 10 block-level code snippets and parsed them to obtain the corresponding statement-level code snippets. Then we will encode the selected code snippets using different models. Finally, we use t-SNE \cite{Maaten2008VisualizingDU} to project the representations of code snippets of different granularities into a two-dimensional space. We marked snippets belonging to the same function with identical colors and used different shapes to denote different granularities: triangles for the function level, squares for the block level, and circles for the statement level. As can be observed from \autoref{fig:MGVisual} (a), the visualization of the original CodeBERT representations shows function-level code snippets clustered together while block-level snippets are dispersed. After applying the MGS$^{3}$ framework, the representations of code snippets from the same function but of different granularities appear more concentrated, and a similar phenomenon is evident in \autoref{fig:MGVisual} (b). This indicates that our framework is capable of achieving superior representations of code snippets across different granularities.

\subsubsection{\textbf{Interpretability Verification}}
To understand whether MGS$^{3}$ can leverage fine-grained representations to enhance coarse-grained search tasks, we attempted to conduct an interpretability analysis on MGS$^{3}$. Specifically, we obtain queries and corresponding function-level code snippets from the COSQA dataset, and determine their importance by calculating the contribution of fine-grained code snippets within functions to the final search score. To achieve this goal, we utilized the Grad-CAM \cite{selvaraju2017grad} method to visualize the gradients of fine-grained code representations relative to the final search score. \autoref{fig:casestudy} displays the visualization results. In the analyzed function, a for loop used to traverse tree nodes and calculate the maximum depth was identified as an important part of understanding the function's semantics and matching it with the query. Observation reveals that this \textit{"for..."} loop made the most significant contribution to the final search score. The visualization results confirm that MGS$^{3}$ can utilize information from code snippets of varying granularities within a function to assist in function-level code search, and demonstrate that MGS$^{3}$ has good interpretability.

\section{Conclusion}
In this paper, we aimed to efficiently leverage self-supervised signals from repositories to enhance model performance across various code search scenarios. We created a dataset, MGCodeSearchNet, which pairs natural language comments with code snippets at different granularities.
We then introduced MGS$^{3}$, a novel multi-granularity code search framework that improves code representation by modeling the hierarchical relationships between code snippets. MGS$^{3}$ also enhances contrastive learning by constructing positive and negative samples from code snippets of varying granularities.
Our experiments on multi-granularity code search benchmarks demonstrated MGS$^{3}$’s superior representation capabilities. Additionally, analytical experiments showed that our method improves the interpretability of pre-trained models in code search tasks.
In future work, we plan to expand the MGCodeSearchNet dataset using a wider range of online repositories and to attempt to combine MGS$^{3}$ with retrieval-augmented generation techniques to achieve better code generation performance.


\begin{acks}
This research was partially supported by grants from the National Natural Science Foundation of China (Grants No. 62337001, 62477044), the Key Technologies R \& D Program of Anhui Province (No. 202423k09020039), the Fundamental Research Funds for the Central Universities and the Iflytek joint research program.
\end{acks}

\bibliographystyle{ACM-Reference-Format}
\bibliography{sample-base}

\appendix

\section{MGCodeSearchNet details}
\label{sec:datadetails}
We have compiled the statistics of the current dataset. The average word count of code comments is 14.9. The average number of lines and tokens (using CodeBERT’s tokenizer) for different granularity of code segments is shown in Table~\ref{tab:level_comparison}.
\begin{table}[h]
    \centering
    \begin{tabular}{lccc}
        \hline
        & \textbf{Function-level} & \textbf{Block-level} & \textbf{Statement-level} \\
        \hline
        \textbf{Lines}  & 14.2 & 5.7 & 1 \\
        \textbf{Tokens} & 122.2 & 51.3 & 15.7 \\
        \hline
    \end{tabular}
    \caption{Comparison of lines and tokens at different levels}
    \label{tab:level_comparison}
\end{table}
%







\section{Benchmark Statictics}
\label{sec:benchdetails}
To validate the performance of MGS$^{3}$ on code search tasks of different granularities, we evaluated it on several existing code search benchmarks of varying granularities:
\begin{itemize}[leftmargin=*]
    \item \textbf{Function level}: CodeSearchNet (CSN) \cite{husain2019codesearchnet, guo2021graphcodebert} collects datasets from six different programming languages from GitHub repositories, which can be used to assess model performance across different programming languages. Adv \cite{lu2021codexglue} is based on the CodeSearch dataset and has normalized method and variable names in the development/test sets to make it more challenging. CoSQA \cite{huang2021cosqa} queries are collected from web search engines, thus better verifying the model's performance in real code search scenarios. XLCoST-FL \cite{zhu2022xlcost} has gathered programming questions along with corresponding programs in seven different programming languages from the GeeksForGeeks\footnote{\url{https://www.geeksforgeeks.org}}.
    \item \textbf{Block level}: SO-DS \cite{heyman2020sods} and StaQC \cite{yao2018staqc} are collected from StackOverflow\footnote{\url{https://www.stackoverflow.com}} questions, which can validate the model's code search performance in programming communities like StackOverflow. XLCoST-BL is a block-level aligned code search dataset obtained through the division of function-level comments in XLCoST. This relies on the well-defined comment specifications on the GeeksForGeeks.
    \item \textbf{Statement level}: CoNaLa \cite{yin2018conala} collects question titles and replies from StackOverflow to construct a code search dataset.
\end{itemize}
The statistical information of the benchmark is shown in~\autoref{tab:datasets}.

\begin{table}[hb]
\caption{The statistics of benchmark datasets.}
\centering
\begin{tabularx}{0.46\textwidth}{lrrrr}
\toprule
Dataset & Granularity & Training & Validation & Test \\
\midrule
CSN-Ruby    & Function & 2.5K      & 1.4K & 1.2K \\
CSN-JS      & Function & 5.8K      & 3.9K & 3.3K \\
CSN-Go      & Function & 16.7K    & 7.3K & 8.1K \\
CSN-Python  & Function & 25.2K      & 13.9K & 14.9K \\
CSN-Java    & Function & 16.4K    & 5.2K & 10.9K \\
CSN-PHP     & Function & 24.1K    & 13.0K & 14.0K \\
Adv  & Function & 28.0K      & 9.6K & 19.2K \\
CoSQA & Function & 19.0K      & 0.5K & 0.5K \\
XLCoST-FL & Function & 9.2K & 0.5K & 0.9K \\ \hline
SO-DS & Block & 14.2K      & 0.9K & 1.1K \\
StaQC & Block & 20.4K      & 2.6K & 2.7K \\
XLCoST-BL & Block & 81.2K      & 3.9K & 7.3K \\ \hline
CoNaLa & Statement & 2.8K      & - & 0.8K \\
\bottomrule
\end{tabularx}
\label{tab:datasets}
\end{table}

\section{Negative Sample Selection}
\label{sec:negalgo}
MGS$^3$ not only uses in-batch negative samples but also adds other code snippets of the same granularity within the same function to expand the negative sample set $c^-$. We consider that these internal negative samples from the same function can be considered more challenging than random samples from blocks and statements of other functions because they share similar function context information. This requires the model to have the ability to distinguish specific functionalities within the same function context. 
It is worth noting that we do not consider code snippets with a nesting relationship within a function as in-function negatives, due to their often close semantic connections. Instead, we choose code snippets that are mutually independent as in-function negatives. 

\begin{figure}[htbp]
\centering
\includegraphics[width=1.0\linewidth]{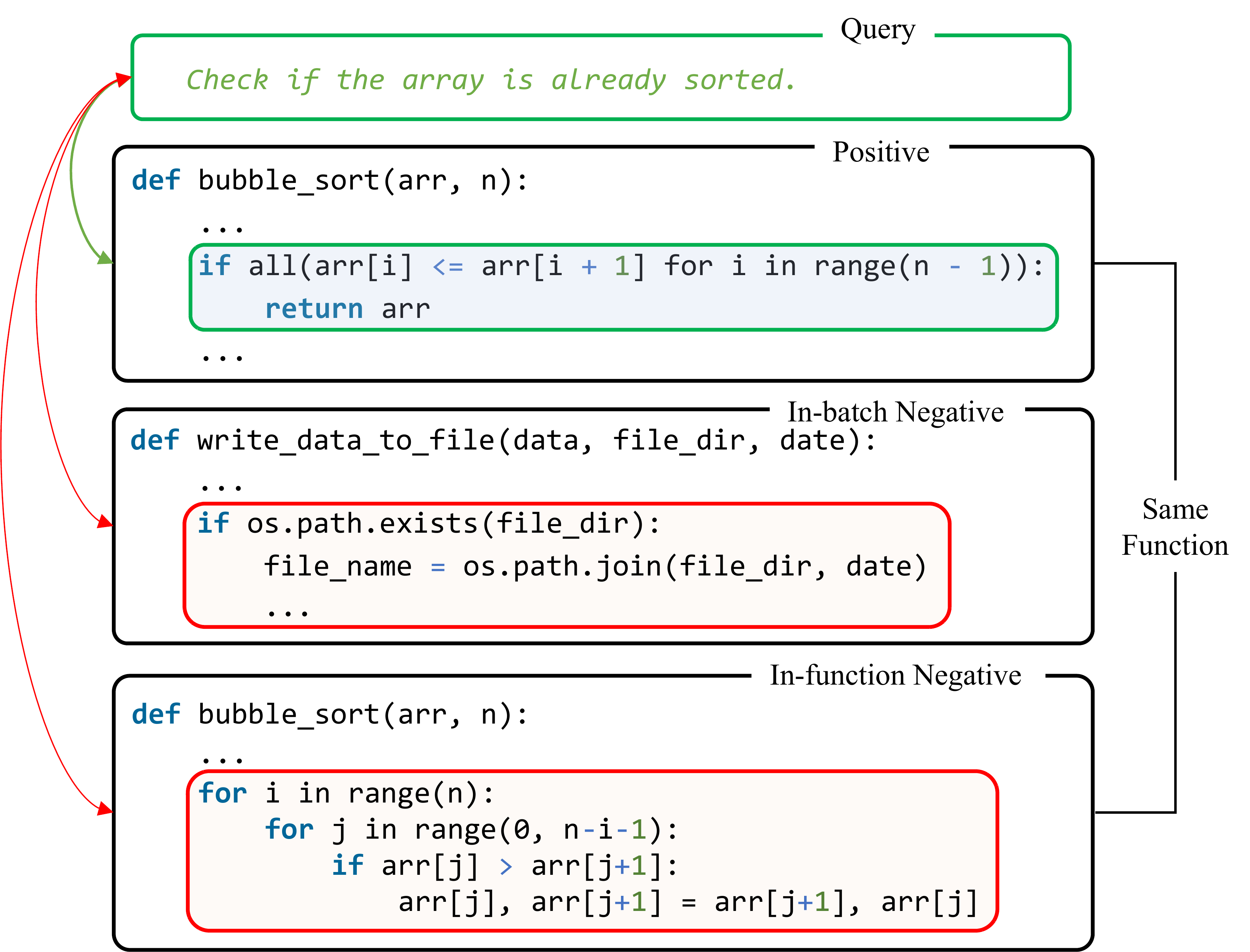}
\caption{We employed methods for acquiring negative samples of fine-grained code snippets, which include in-batch negatives and in-function negatives, taking block-level code snippets as an example.}
\label{fig:negative}
\end{figure}






\begin{table*}[tb]
\caption{In benchmarks of code search at different granularities, we compare the fine-tuning performance of pre-trained code representation models with the performance under a fine-tuning setting using MGS$^{3}$.}
\begin{tabular}{cccccccccc}
\hline
\multirow{2}{*}{Granularity} & \multirow{2}{*}{Dataset} & \multicolumn{2}{c}{CodeBERT} & \multicolumn{2}{c}{GraphCodeBERT} & \multicolumn{2}{c}{UniXCoder} & \multicolumn{2}{c}{CodeT5+} \\ \cline{3-10} 
                             &                          & Fine-tuning      & w/ MGS$^{3}$      & Fine-tuning         & w/ MGS$^{3}$        & Fine-tuning       & w/ MGS$^{3}$      & Fine-tuning      & w/ MGS$^{3}$     \\ \hline
\multirow{8}{*}{Function-level}    & CSN-Ruby                  &  67.9            & \textbf{70.7}            & 70.3                & \textbf{72.5}              & 74.0              & \textbf{75.7}            & 74.9             & \textbf{76.0}           \\
                             & CSN-JS                    & 62.0             & \textbf{64.0}            & 64.4                & \textbf{66.6}              & 68.4              & \textbf{69.1}            & 69.2             & \textbf{69.8}           \\
                             & CSN-Go                    & 88.2             & \textbf{89.9}            & 89.7                & \textbf{91.3}               & 91.5              & \textbf{92.1}            & 91.0             & \textbf{91.8}           \\
                             & CSN-Python                & 67.2             & \textbf{69.4}            & 69.2                & \textbf{70.7}              & 72.0              & \textbf{73.6}            & 71.9             & \textbf{72.8}           \\
                             & CSN-Java                  & 67.6             & \textbf{69.5}            & 69.1                & \textbf{71.3}              & 72.6              & \textbf{74.8}            & 72.4             &\textbf{73.6}           \\
                             & CSN-PHP                   & 62.8             & \textbf{64.7}            & 64.9                & \textbf{66.5}              & 67.6              & \textbf{69.0}            & 68.2             & \textbf{69.3}           \\
                             & Adv                      & 27.2             & \textbf{29.4}            & 35.2                & \textbf{37.2}              & 41.3              & \textbf{42.5}            &  43.3             & \textbf{44.7}           \\
                             & CoSQA                    & 64.7             & \textbf{66.0}            & 67.5                & \textbf{68.8}              & 70.1              & \textbf{71.3}            & 72.7             & \textbf{73.9}           \\ 
                             & XLCoST-FL                    & 58.0             & \textbf{60.7}            & 59.4                & \textbf{62.2}              & 59.1              & \textbf{61.9}            & 63.3             & \textbf{65.1}           \\ \hline                             
\multirow{2}{*}{Block-level}       & SO-DS                    & 23.1             & \textbf{27.3}            & 25.3                & \textbf{27.9}              & 23.6              & \textbf{27.3}            & 26.1             & \textbf{29.5}           \\
                             & StaQC                    & 23.4             & \textbf{27.6}            & 23.8                & \textbf{28.7}              & 23.1              & \textbf{28.2}            & 25.7             & \textbf{29.4}           \\ 
                             & XLCoST-BL                    & 30.9             & \textbf{38.9}            & 31.6                &  \textbf{39.5}             & 30.2              & \textbf{38.0}            & 35.8             & \textbf{40.0}           \\ \hline                             
Statement-level                    & CoNaLa                   & 20.9             & \textbf{25.0}            & 23.5                & \textbf{26.9}              & 20.4              & \textbf{25.1}            & 22.8             & \textbf{27.1}           \\ \hline
\end{tabular}
\label{tab:extraExperi}
\end{table*}

\begin{table*}[tb]
\caption{The results of ablation experiments conducted on different granularity code search benchmarks.}
\begin{tabular}{cl|ccc|ccc|c}
\hline
\multicolumn{2}{l|}{\multirow{2}{*}{}}                             & \multicolumn{3}{c|}{Function-level}                                                  & \multicolumn{3}{c|}{Block-level}                                                       & \multicolumn{1}{c}{Statement-level} \\ \cline{3-9} 
\multicolumn{2}{l|}{}                                              & \multicolumn{1}{c}{Adv} & \multicolumn{1}{c}{CoSQA} & \multicolumn{1}{c|}{XLCoST-FL} & \multicolumn{1}{c}{SO-DS} & \multicolumn{1}{c}{StaQC} & \multicolumn{1}{c|}{XLCoST-BL} & \multicolumn{1}{c}{CoNaLa}          \\ \hline
\multicolumn{2}{c|}{CodeBERT+MGS3}                                 & 25.6                       & 36.6                         & 25.9                              & 16.4                         & 14.7                         & 20.7                              & 12.8                                   \\ \hline
\multicolumn{1}{c|}{\multirow{2}{*}{Granularity}} & w/o $\mathcal{L}_b$       & 23.2                       & 34.2                         & 23.6                              & 9.5                         & 8.8                         & 14.8                              & 10.7                                   \\
\multicolumn{1}{c|}{}                             & w/o $\mathcal{L}_s$       & 25.5                       & 36.7                         & 26.3                              & 13.4                         & 13.0                         & 16.9                              & 9.8                                   \\ \hline
\multicolumn{1}{c|}{\multirow{3}{*}{Module}}      & w/o HMGR        & 23.8                       & 35.2                         & 23.9                              & 15.4                         & 14.6                         & 19.7                              & 12.8                                   \\
\multicolumn{1}{c|}{}                             & w/o Pos-sample      & 24.6                       & 35.7                         & 24.8                              & 14.2                         & 13.2                         & 18.1                              & 11.5                                   \\
\multicolumn{1}{c|}{}                             & w/o Neg-sample & 24.3                       & 35.1                         & 24.5                             &  12.9                        & 12.7                         & 17.3                              & 10.7                                   \\ \hline
\end{tabular}
\label{tab:ablation}
\end{table*}
\section{Extra of Experiment}

\subsection{Results of the Fine-tuning Experiment.}
We present the results of the fine-tuning in~\autoref{tab:extraExperi}. By observing the experimental results on the CodeSearchNet dataset across various programming languages, it can be seen that by applying MGS$^3$ method, significant improvements can be achieved in all programming languages, which demonstrates the effectiveness of the MGS$^3$.

\subsection{Ablation Study}
\label{sec:ablation}
We attempted to present the complete ablation study experiment in~\autoref{tab:ablation}. We can observe that different granular data sets exhibit relatively similar changes for different ablation experiment settings. In the block-level experimental results, we observed a significant performance drop in the SO-DS benchmark after deleting $\mathcal{L}_s$. We speculate that this is because the benchmark comes from StackOverflow, and during the collection process, some statement-level code snippets were mixed in.

\subsection{Parameter Sensitivity Experiments}
\label{sec:paraSensty}
For the weights $\alpha$ and $\beta$ in the loss functions of different granularities, we conducted parameter sensitivity experiments using the CodeBERT model on Adv, StaQC, and CoNaLa. Table 6 shows the performance of $\alpha$ and $\beta$ at 0.2, 0.4, 0.6, 0.8, and 1. When adjusting one of the parameters, we set the other to 0 to facilitate the observation of results. 


\begin{table}[htb]
\caption{Results of sensitivity experiments on $\alpha$ and $\beta$.}
\begin{tabular}{cccc|cccc}
\hline
$\alpha$ & Adv  & StaQC & CoNaLa & $\beta$ & Adv  & StaQC & CoNaLa \\ \hline
0.2      & 24.2 & 4.0   & 3.5    & 0.2     & \textbf{24.2} & 3.2   & 5.7    \\
0.4      & 24.5 & 7.8   & 5.9    & 0.4     & 24.0 & 7.1   & 8.9    \\
0.6      & 25.1 & 10.2  & 7.4    & 0.6     & 23.7 & \textbf{9.0}   & 10.0    \\
0.8      & \textbf{25.7} & 11.6  & 8.5    & 0.8     & 23.0 & 8.7   & 10.2   \\
1.0      & 25.5 & \textbf{13.0}  & \textbf{9.8}    & 1.0     & 23.2 & 8.8   & \textbf{10.6}   \\ \hline
\end{tabular}
\label{tab:sensity}
\end{table}

\end{document}